\documentclass[preprint,review,12pt]{elsarticle}

\usepackage{amssymb}
\usepackage{subcaption}
\usepackage{graphicx}
\usepackage{tabularx}
\usepackage{multirow}
\usepackage{makecell}

\usepackage{lineno}
\usepackage{color}
\usepackage{array}
\usepackage{amsmath}

\newcommand{\pT}{$p_{\rm{T}}$}
\newcommand{\pTtwo}{$\langle p^{2}_{\rm{T}} \rangle$}

\newcommand{\raa}{$R_{\mathrm{AA}}$}

\journal{Physics Letters B}

\biboptions{sort&compress}
\begin{document}

\begin{frontmatter}



\title{Measurement of inclusive $J/\psi$ production in Au+Au collisions at $\sqrt{s_\mathrm{NN}} = 54.4$ GeV at STAR}


\author{
B.~E.~Aboona$^{58}$,
J.~Adam$^{17}$,
L.~Adamczyk$^{3}$,
I.~Aggarwal$^{45}$,
M.~M.~Aggarwal$^{45}$,
Z.~Ahammed$^{66}$,
A.~K.~Alshammri$^{33}$,
E.~C.~Aschenauer$^{7}$,
S.~Aslam$^{22}$,
J.~Atchison$^{2}$
% ...
}

\address{\rm{(STAR Collaboration)}}

% ...
\address{$^{2}$Abilene Christian University, Abilene, Texas   79699}
\address{$^{3}$AGH University of Krakow, FPACS, Cracow 30-059, Poland}
% ...
\address{$^{7}$Brookhaven National Laboratory, Upton, New York 11973}
% ...
\address{$^{17}$Czech Technical University in Prague, FNSPE, Prague 115 19, Czech Republic}
% ...
\address{$^{22}$Fudan University, Shanghai, 200433 }
% ...
\address{$^{33}$Kent State University, Kent, Ohio 44242}
% ...
\address{$^{45}$Panjab University, Chandigarh 160014, India}
% ...
\address{$^{58}$Texas A\&M University, College Station, Texas 77843}
% ...
\address{$^{66}$Variable Energy Cyclotron Centre, Kolkata 700064, India}
% ...

\begin{abstract}
This article presents measurements of inclusive $J/\psi$ production at midrapidity ($\left|y\right| <$ 1.0) in Au+Au collisions at $\sqrt{s_\mathrm{NN}} = 54.4$ GeV with the STAR detector at the Relativistic Heavy Ion Collider. A suppression of the $J/\psi$ yield, quantified using the nuclear modification factors ($R_{\mathrm{AA}}$, $R_{\mathrm{CP}}$), is observed with respect to the scaled production in $p$+$p$ collisions. The dependence of $R_{\mathrm{AA}}$ on collision centrality and $J/\psi$ transverse momentum is measured with improved precision compared to previous measurements at 39 and 62.4 GeV, while the centrality dependence of $R_{\mathrm{CP}}$ is measured and compared to the same results at 39, 62.4, and 200 GeV. In central collisions, no significant collision energy dependence of $R_{\mathrm{AA}}$ is found within uncertainties for collision energies between 17.3 and 200 GeV. Two transport model calculations that include dissociation and regeneration contributions are consistent with the experimental results within uncertainties. Although no significant collision energy dependence of the $J/\psi$ suppression in high energy heavy-ion collisions up to $\sqrt{s_\mathrm{NN}} = 200$ GeV is observed within uncertainties, the newly measured results at 54.4 GeV Au+Au collisions provide additional constraints on theoretical calculations of the hot medium evolution and cold nuclear matter effects.
\end{abstract}




\end{frontmatter}


\section{Introduction}
\label{sec1}
A deconfined state of partonic matter, predicted by Quantum Chromodynamics (QCD) and called the Quark-Gluon Plasma (QGP), is produced in ultrarelativistic heavy-ion collisions at the Super Proton Synchrotron (SPS)~\cite{Heinz:2000bk}, the Relativistic Heavy Ion Collider (RHIC)~\cite{ADAMS2005102,ARSENE20051,PHENIX:2004vcz,PHOBOS:2004zne} and the Large Hadron Collider (LHC)~\cite{ALICE:2022wpn}. Understanding the characteristics of the QGP is one of the main research goals of current high-energy heavy-ion experiments. Among the various probes used to study the QGP properties, quarkonia play a unique role. Quarkonia are bound states of heavy quarks and their anti-quarks, whose masses are significantly larger than the QCD scale. Their production yields in heavy-ion collisions are subject to modifications in the presence of the QGP. The hot medium effects on quarkonia include dissociation due to static color screening of the potential between the heavy quark pair~\cite{MATSUI1986416}, dynamical color screening or collisional dissociation caused by interactions with medium constituents~\cite{PhysRevC.53.3051,PhysRevD.100.014008,PhysRevC.87.044905}, and regeneration from deconfined heavy quarks and anti-quarks~\cite{BRAUNMUNZINGER2000196,PhysRevLett.92.212301}. Due to the presence of nuclei in the collisions, there are also unavoidable modifications from Cold Nuclear Matter (CNM) effects. These effects arise from changes in parton distribution functions in nuclei compared to those in free nucleons, energy loss in the colliding nuclei, the Cronin effect, nuclear absorption, and other final state effects such as dissociation by co-movers~\cite{PhysRevC.84.044911,NAGLE199921,PhysRevD.11.3105,VOGT2002539,FERREIRO201598}.

The $J/\psi$ meson~\cite{PhysRevLett.33.1404,PhysRevLett.33.1406,STAR:2018smh,PHENIX:2009ghc,Smith:2022pro}, bound state of a pair of charm and anti-charm quarks, is the most abundantly produced quarkonium state accessible experimentally. In heavy-ion collisions, the inclusive $J/\psi$ production includes four main sources: $J/\psi$ mesons produced directly during the initial partonic scatterings and via regeneration, those from decays of excited charmonium states, such as $\chi_{c}$ and $\psi$(2S), as well as those from decays of long-lived $b$-hadrons (called non-prompt $J/\psi$). Precision measurements of $J/\psi$ production yields at mid-rapidity have been achieved in Pb+Pb collisions at center-of-mass energies per nucleon-nucleon pair ($\sqrt{s_\mathrm{NN}}$) of 17.3 GeV at the SPS~\cite{Kluberg:2005yh,200028}, 2.76 and 5.02 TeV at the LHC~\cite{2014314,ALICE:2023gco}, as well as in Au+Au collisions at $\sqrt{s_\mathrm{NN}}$ = 39, 62.4 and 200 GeV at RHIC~\cite{PhysRevLett.98.232301,STAR:2012wnc,STAR:2013eve,201713,2019134917}. 

Significant suppression of J/$\psi$ production in heavy-ion collisions compared to that in $p$+$p$ collisions of the same energy is observed at SPS and RHIC~\cite{Kluberg:2005yh,200028,PhysRevLett.98.232301,STAR:2012wnc,STAR:2013eve,201713,2019134917}. The level of suppression is found to be similar, although the temperature and energy density of the produced medium are significantly different due to an order of magnitude difference in $\sqrt{s_\mathrm{NN}}$~\cite{PhysRevLett.98.232301}. The similarity of the suppression is interpreted as the interplay of the energy-dependent dissociation in the QGP and CNM effects, with the regeneration contribution only becoming significant at RHIC. The importance of the regeneration process is corroborated by the observation of decreasing $J/\psi$ suppression from the top RHIC energy to LHC energies~\cite{2014314,ALICE:2015nvt,BAI2021121769}, owning to enhanced regeneration contribution driven by the larger charm quark cross section at higher collision energy.


There are several theoretical models that employ different dynamic processes, such as $J/\psi$ mesons considered to be dissociated and regenerated continuously during the medium evolution~\cite{PhysRevLett.97.232301,PhysRevC.89.054911,PhysRevC.82.064905,ZHAO2011114} or to be completely melted above the dissociation temperature and then regenerated at the chemical freeze-out \cite{BRAUNMUNZINGER2000196,Andronic:2017pug}. These models can qualitatively describe the experimental measurements. However, the uncertainties in theoretical calculations remain substantial. This is because the underlying mechanisms responsible for the hot medium and CNM effects are not well understood. The model parameters, such as the dissociation temperatures of quarkonium states, the temperature profile of the medium, the heavy quark production cross section and the dynamic evolution of heavy quarks in the medium, are also poorly constrained, which further contributes to the large uncertainties in theoretical models. All these aspects strongly depend on the collision energy, and thus a fine collision energy scan of quarkonium production in heavy-ion collisions can provide stringent constraints on theoretical calculations, which are essential for inferring the properties of the QGP from quarkonium measurements.

The RHIC beam energy scan program, mainly designed for the exploration of the QCD phase structure and to search for the critical end point~\cite{Chen:2024aom}, enables a collision energy scan of $J/\psi$ production between the SPS energy and the top RHIC energy. The STAR and PHENIX experiments have measured $J/\psi$ suppression in Au+Au collisions at $\sqrt{s_\mathrm{NN}}=$ 39 and 62.4 GeV for mid- and forward rapidities, respectively~\cite{201713,PHENIX:2012xtg}, and no significant energy dependence is observed. However, the experimental uncertainties are large due to the limited statistics available. In 2017, the STAR collaboration recorded a significantly larger sample of Au+Au collisions at $\sqrt{s_\mathrm{NN}}=$ 54.4 GeV compared to those at 39 and 62.4 GeV, with about a factor of six more events. Measurements of $J/\psi$ production at a new energy with improved precision will help to further constrain both the cold and hot medium effects on the $J/\psi$ production in heavy-ion collisions.


In this paper, measurements of the transverse momentum ($p_{\mathrm{T}}$) and centrality dependence of the inclusive $J/\psi$ production at mid-rapidity ($\left | y \right | <1$) through the dielectron decay channel in Au+Au collisions at $\sqrt{s_\mathrm{NN}} = 54.4$ GeV are presented. \pTtwo\ of inclusive $J/\psi$ meson is extracted as a function of the collision centrality. The collision energy dependence of the inclusive $J/\psi$ production is further studied by comparing the new measurements with the published data and theoretical calculations.

\begin{table*}[htbp]
\centering
\begin{tabular}{|c | c | c|}
\hline
    Track \pT\ & PID detectors & Electron PID cuts  \\
\hline
    \pT\ $\leq 1.0$ GeV/$c$ & TPC, TOF & {\small\makecell{$\left|1/\beta - 1 \right|<0.025$; \\ $p\ \leq\ 0.8$ GeV/c: $3\times p - 3.15\ \textless\ n\sigma_{e}\ \textless\ 2$,\\ $p\ \textgreater\ 0.8$ GeV/c: $-0.75\ \textless\ n\sigma_{e}\ \textless\ 2$}}\\
\hline
    \multirow{7}{*}{\pT\ $ > 1.0$ GeV/$c$} & \makecell{TPC, TOF \\(not matched to BEMC)} & {\small\makecell{$\left|1/\beta - 1 \right|<0.025$; \\  $-0.75\ \textless\ n\sigma_{e}\ \textless\ 2$ } } \\
\cline{2-3}
      & \multirow{3}{*}{\makecell{TPC, BEMC \\(not matched to TOF)}} & \multirow{3}{*}{{\small\makecell{$-1\ \textless\ n\sigma_{e}\ \textless\ 2$  ; \\ $0.5\ \textless\ E_{0}/p \ \textless\ 1.5$} }}\\
      & & \\
      & & \\
\cline{2-3}
     & \makecell{TPC, TOF, and BEMC} & {\small\makecell{$\left|1/\beta - 1 \right|<0.025$;\\  $-1.5\ \textless\ n\sigma_{e}\ \textless\ 2$; \\$0.5\ \textless\ E_{0}/p\ \textless\ 1.5$ } } \\
\hline
   
\end{tabular}
\caption{List of detectors used for electron PID and corresponding cuts in different $p_{\mathrm{T}}$ intervals.}\label{table1}
\label{table1}
\end{table*}

\section{Experiment and analysis}
\label{sec2}

The STAR experiment is a general-purpose detector built to study the formation and characteristics of the QGP. It covers the entire azimuth within the pseudorapidity interval of $\left|\eta\right| <$ 1. The dataset utilized in this analysis contains $1.3 \times 10^{9}$ minimum-bias events for Au+Au collisions at $\sqrt{s_\mathrm{NN}} = 54.4$ GeV collected by the STAR experiment in 2017. The minimum-bias trigger requires coincident signals in the two Zero-Degree Calorimeters (ZDCs)~\cite{Judd:2018zbg,Adler:2000bd}, or the two Vertex Position Detectors (VPDs)~\cite{LLOPE201423}. Charged particles produced in the collisions are tracked and reconstructed in the Time Projection Chamber (TPC)~\cite{Anderson:2003ur}, which is immersed in a solenoidal magnetic field of 0.5 T. The event vertex determined using the tracks in the TPC is required to be within $\pm$60 cm (2 cm) from the nominal center of the STAR detector along the beam (radial) direction. To reject pileup events, the reconstructed vertex position along the beam direction is further required to be within 3 cm of the vertex position reconstructed based on coincident signals in the VPDs, which are fast detectors and thus resilient to pileup effects. To avoid trigger detector inefficiency for events with low multiplicities, only data from the 0-60\% most central collisions are analyzed. The total number of events used after event selection is 593 million.

The collision centrality is determined by matching the multiplicity distribution of charged tracks ($dN_{\mathrm{ch}}/d\eta$) in data to that from the Monte Carlo Glauber model~\cite{PhysRevC.79.034909}. The $dN_{\mathrm{ch}}/d\eta$ is obtained by counting the number of charged tracks within $\left|\eta\right| <$ 0.5 and is corrected for vertex position and luminosity dependences, but not for the tracking efficiency. The average number of binary collisions ($\langle N_\mathrm{coll} \rangle$) and number of participants ($\langle N_\mathrm{part} \rangle$) for each centrality bin are determined using the Glauber model. 

The $J/\psi \to e^{+}e^{-}$ decay channel is used to reconstruct $J/\psi$ candidates in this analysis. To assure high quality, accepted tracks in the TPC are required to have at least 20 space points used for their reconstruction. The tracks' distance of closest approach to the event vertex needs to be less than 1 cm. Furthermore, the ratio of the number of TPC space points used for track reconstruction to the maximum possible number along the track trajectory should be larger than 0.52 to remove split tracks.

The main detectors used for electron identification are the TPC, the Time-of-Flight (TOF) detector~\cite{LLOPE2012S110}, and the Barrel Electromagnetic Calorimeter (BEMC)~\cite{STAR:2002ymp}. Unless specified otherwise, the term ``electron" denotes both electrons and positrons in the following. The TPC provides particle identification (PID) through measurement of the specific energy loss ($\langle dE/dx \rangle$). The variable used is defined as: 
\begin{equation}
\setlength{\abovedisplayskip}{6pt}
\setlength{\belowdisplayskip}{6pt}
\begin{aligned}
 n\sigma_{\mathrm{e}} = \frac{\ln (dE/dx)_{\rm{measured}} - \ln (dE/dx)^{\mathrm{e}}_{\rm{theory}}}{R_{\ln(dE/dx)}},
\end{aligned}
\end{equation}
where $(dE/dx)_{\rm{measured}}$ is the measured energy loss, $(dE/dx)^{\mathrm{e}}_{\rm{theory}}$ is the predicted energy loss for an electron based on the Bichsel formalism~\cite{BICHSEL2006154} and $R_{\ln (dE/dx)}$ is the resolution of $\ln (dE/dx)$. Consequently, the $n\sigma_{\mathrm{e}}$ distribution for electrons is expected to follow a Gaussian distribution with a mean of zero and width of one. To ensure good resolution, the number of TPC space points used to calculate $dE/dx$ is required to be no less than 15. In combination with the path length measured in the TPC, the TOF provides the velocity information ($\beta=v/c$) by measuring a particle's flight time, where $c$ is the speed of light. Electrons can be separated from hadrons due to their larger velocities at a given momentum. Furthermore, the BEMC is used to suppress hadron contamination via the measurement of deposited energies by incident particles and comparison to their momenta measured in the TPC. A TPC track is matched to a BEMC cluster geometrically by projecting to the BEMC surface, and the variable $E_{0}/p$, where $p$ is the particle momentum measured in the TPC and $E_{0}$ is the energy deposition of the most energetic tower in the matched BEMC cluster, is used for PID \cite{PhysRevC.81.064904}. For electrons, the $E_{0}/p$ is expected to be peaked around one, while for hadrons it is considerably smaller than one.

Table \ref{table1} lists the combinations of detectors and specific cut values used for electron identification in different \pT\ ranges. For tracks with $p_{\mathrm{T}} \leq 1$ GeV/$c$, a momentum-dependent $n\sigma_{\mathrm{e}}$ cut is applied, along with the requirement $\left|1/\beta - 1 \right|<0.025$. For $p_{\mathrm{T}} > 1$ GeV/$c$, different combinations of PID selections are used, depending on the availability of TOF and BEMC information, in order to maximize the hadron rejection power while maintaining a high identification efficiency. When the track is only matched to TOF, the selection criteria of $\left|1/\beta - 1 \right|<0.025$ and $-0.75 < n\sigma_{\mathrm{e}} < 2$ are applied utilizing both TOF and TPC information. If a track leaves a signal in the BEMC but not in the TOF, the BEMC cut of $0.5 < E_{0}/p < 1.5$ is utilized in conjunction with the TPC cut of $-1 < n\sigma_{\mathrm{e}} < 2$. The third scenario is that the PID information from TPC, TOF and BEMC can all be used, and the corresponding electron selection cuts are: $\left|1/\beta - 1 \right|<0.025$, $-1.5 < n\sigma_{\mathrm{e}} < 2$ and $0.5 < E_{0}/p < 1.5$. It is worth noting that the $n\sigma_{\mathrm{e}}$ cut is asymmetric with a tighter and lower boundary on the negative side. This is because charged pions are the most abundant hadron species that contaminate the electron sample and their $n\sigma_{\mathrm{e}}$ values are negative. The lower boundary of the $n\sigma_{\mathrm{e}}$ cut also changes with $p_{\mathrm{T}}$ and the availability of TOF and BEMC information, while the upper boundary is kept at 2.

\begin{figure}[htbp]
\centering
\includegraphics[width=0.9\columnwidth,clip]{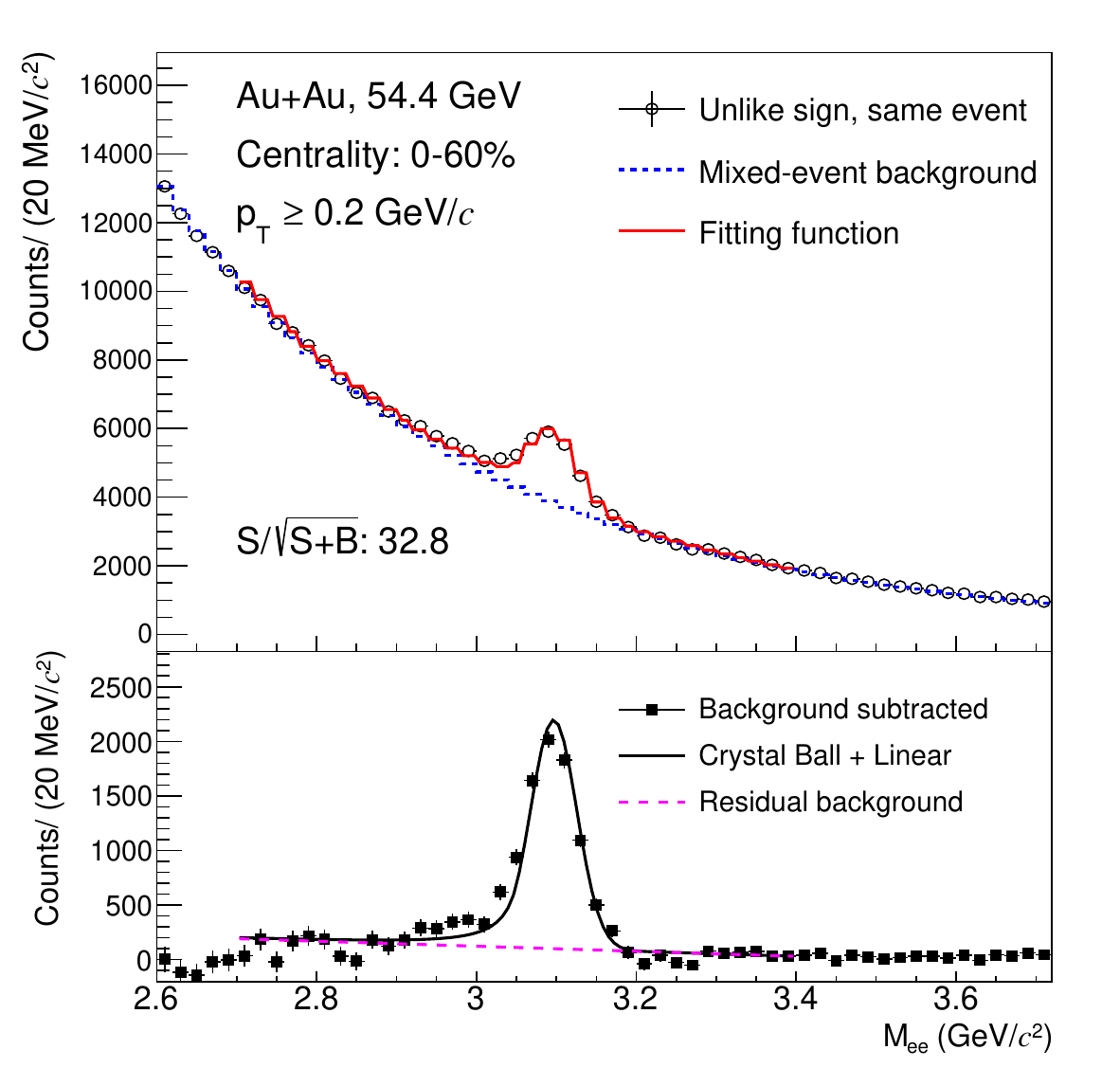}
\caption{(top) Invariant mass distributions for unlike-sign pairs from same events and mixed events in 0-60\% central Au+Au collisions at $\sqrt{s_\mathrm{NN}} = 54.4$ GeV. (bottom) Difference of unlike-sign distributions between same and mixed events, overlaid with the fit result of the $J/\psi$ signal plus the residual background.}
\label{fig-1}
\end{figure}

Figure~\ref{fig-1} shows the invariant mass ($M_{\mathrm{ee}}$) distribution of $e^{+}e^{-}$ pairs in 0-60$\%$ Au+Au collisions at $\sqrt{s_\mathrm{NN}} = 54.4$ GeV. A $p_{\rm{T}}>0.2$ GeV/$c$ cut is applied to exclude coherent photon-induced $J/\psi$ production \cite{PhysRevLett.116.222301,PhysRevLett.123.132302,PhysRevC.97.044910}.
The raw yield of the $J/\psi$ meson is obtained by counting the $e^{+}e^{-}$ pairs in the mass range of 2.9-3.2 GeV/$c^{2}$, where a clear $J/\psi$ peak is seen around $M_{\mathrm{ee}} = 3.09$ GeV/$c^{2}$, and subtracting the combinatorial and residual background within the mass window. The residual background, originating from correlated heavy-flavor decays and Drell-Yan processes, is estimated by fitting the same-event unlike-sign distribution using the unbinned maximum likelihood method with the following components: i) the $J/\psi$ signal shape described by a Crystal-Ball function~\cite{Skwarnicki:1986xj,Oreglia:1980cs} whose width is constrained by detector simulations tuned to reproduce realistic momentum resolution; ii) the combinatorial background estimated using the normalized unlike-sign $e^+e^-$ pairs from mixed events \cite{PhysRevC.81.034911}; iii) a first-order polynomial function describing the residual background. The mixed-event unlike-sign distribution is normalized according to the ratio of like-sign distributions between same and mixed events.  The fitting result is shown as the red solid line in the upper panel of Fig.~\ref{fig-1}. Also shown in Fig.~\ref{fig-1} is the combinatorial background subtracted unlike-sign distribution, overlaid with the fit results of the $J/\psi$ signal plus the residual background, in the bottom panel. The obtained $J/\psi$ raw counts is 9085 for $p_{\rm{T}}>0.2$ GeV/$c$ within 0-60\% centrality class, and the corresponding significance, calculated as $S/\sqrt{S+B}$ where $S$ and $B$ are signal and background counts, is 32.8. The efficiency of counting raw $J/\psi$ yields within the mass range of 2.9 - 3.2 GeV/$c^{2}$ is obtained based on the fitted Crystal-Ball function, which is around 92\% with a mild centrality and $p_{\rm{T}}$ dependence.

The electron detection efficiency consists of the TPC tracking efficiency, TOF and BEMC matching efficiencies and PID cut efficiencies. The embedding technique is used to evaluate the TPC tracking efficiency as well as the BEMC matching and PID efficiencies \cite{PhysRevC.79.034909}. A sub-sample of minimum-bias events, which cover the entire period of data taking, is processed using the standard STAR reconstruction procedure after being embedded with the simulated electron tracks propagated through the GEANT3 simulation \cite{PhysRevC.79.034909} of the STAR detector. On the other hand, the TOF matching efficiency as well as the TPC and TOF PID cut efficiencies are evaluated using a pure electron sample in data, originating from photon conversions and light meson Dalitz decays~\cite{PhysRevLett.113.022301}. For an electron track with $p_{\rm{T}}$ around 1.5 GeV/$c$, the TPC track efficiency is about 83\%, the TOF matching efficiency is about 7\% and the BEMC matching efficiency is about 86\% in 0-60\% centrality. The $J/\psi$ reconstruction efficiency is determined by convoluting the electron detection efficiency as a function of \pT, $\eta$, and $\phi$ with the $J/\psi$ decay kinematics.

Systematic uncertainties include those in signal extraction, TPC tracking efficiency, TOF and BEMC matching efficiency estimations and electron identification. For the signal extraction, the invariant mass range is varied when obtaining the normalization factor for the unlike-sign distribution in mixed events. The raw $J/\psi$ counts are obtained directly from the fit function instead of the default bin-counting method. Additionally, different fitting ranges for estimating the background contribution are tried. The shape of the residual background is changed from the default first-order polynomial function to zeroth-order or second-order function. Finally, the width of the Crystal-Ball function is varied taking into account the uncertainties in tuning the simulation to match momentum resolution in data. The maximum deviation of all the variations to the default is taken as the uncertainty. For the TPC tracking efficiency, track quality cuts are varied simultaneously in data analysis and in extracting the tracking efficiency from embedding to evaluate the uncertainties. An additional $5\%$ systematic uncertainty is included for each single electron tracking efficiency to account for discrepancies arising from imperfect simulation when compared to the data~\cite{PhysRevLett.130.202301}. In terms of the electron identification, the uncertainties on $n\sigma_{\mathrm{e}}$ and $E_{0}/p$ cut efficiencies are estimated by changing these cuts and checking the impact on the corrected $J/\psi$ yields. The default 1/$\beta$ cut efficiency is calculated via counting the fraction of the electron 1/$\beta$ distribution within $\left|1/\beta - 1 \right|<0.025$. The difference between the default method and that based on a Gaussian fit to the 1/$\beta$ distribution is taken as the uncertainty. The uncertainty in the TOF matching efficiency is evaluated by comparing efficiency differences when using different cuts to select the electron sample. On the other hand, for the BEMC matching efficiency, the uncertainty originates from the difference between utilizing detector simulation and data-driven approach. All the aforementioned uncertainties are listed in Table \ref{table2}. Global uncertainties, shown as bands at unity in the figures, are from the uncertainties in $\left\langle N_{\rm{coll}}\right\rangle$ and the $J/\psi$ production cross section in $p$+$p$ collisions.

\begin{table}[htbp]
\centering
\begin{tabular}{|c c c c c|}
\hline
    Sources & 0-60\% & 0-20\% & 20-40\% & 40-60\% \\
\hline
    Signal extraction & 5.1\% & 5.7\% & 4.1\% & 3.0\% \\
    TPC tracking & 10.6\% & 10.6\% & 10.6\% & 10.6\% \\
    TOF matching & 0.2\% & 0.1\% & 0.3\% & 0.3\% \\
    BEMC matching & 0.4\% & 0.7\% & 0.7\% & 0.6\% \\
    1/$\beta$ cut & 5.5\% & 5.9\% & 5.2\% & 4.6\% \\
    $n\sigma_{\mathrm{e}}$ cut & 2.1\% & 3.5\% & 2.3\% & 4.1\% \\
    $E_{0}/p$ cut & 5.0\% & 4.8\% & 6.3\% & 6.1\% \\
\hline
    Total & 14\% & 15\% & 14\% & 14\% \\
\hline
   
\end{tabular}
\caption{Individual and total systematic uncertainties for $J/\psi$ with \pT\ $>$ 0.2 GeV/$c$ in different centrality classes.}\label{table2}
\label{table2}
\end{table}

\section{Results and discussions}
\label{sec3}
\begin{figure}[htbp]
\centering
\includegraphics[width=0.9\columnwidth,clip]{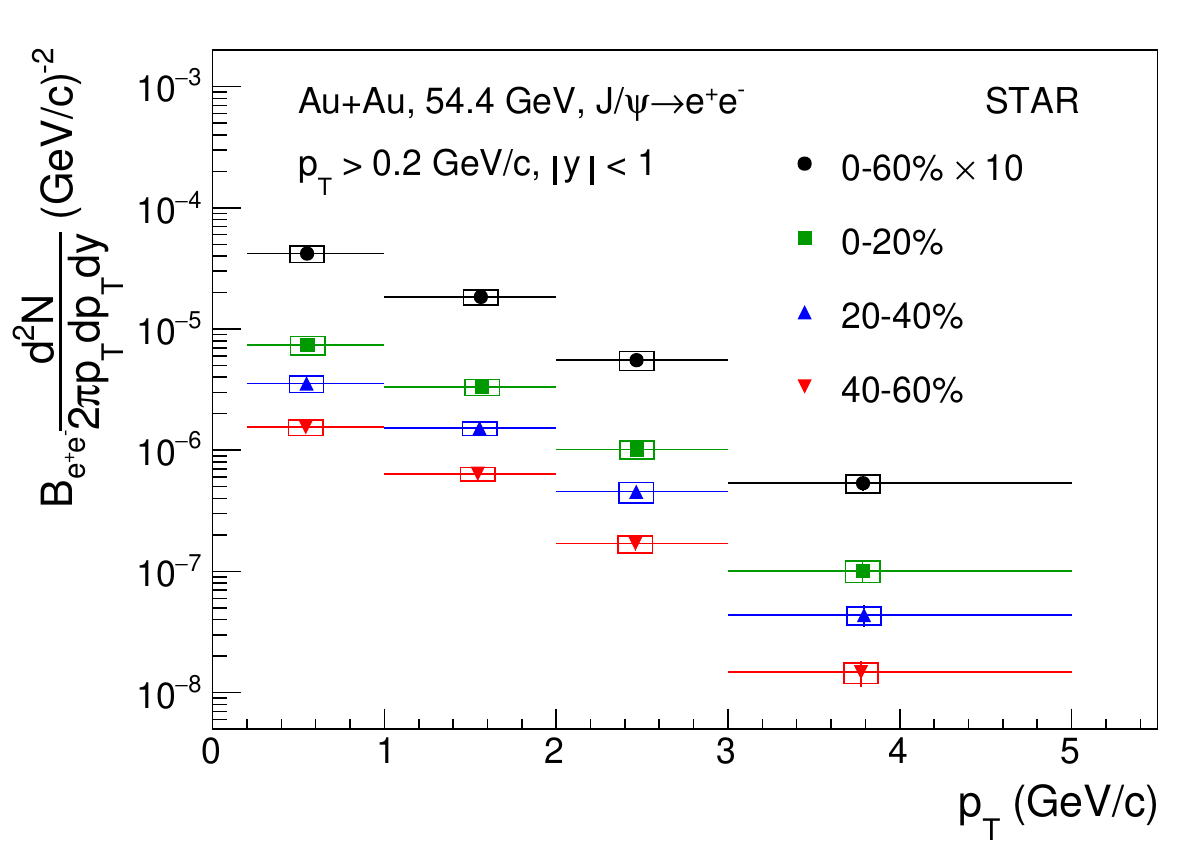}
\caption{Inclusive $J/\psi$ invariant yields as a function of $p_{\mathrm{T}}$ at mid-rapidity ($\left | y \right | <1$) in Au+Au collisions at $\sqrt{s_\mathrm{NN}} = 54.4$ GeV in different centralities. The vertical error bars represent the statistical uncertainties, while boxes represent the systematic uncertainties. The horizontal bars depict the $p_{\mathrm{T}}$ binning. Data points for 0-60\% centrality are scaled up by a factor of ten for clarity.}
\label{fig-2}
\end{figure}

The efficiency and acceptance corrected invariant yields of inclusive $J/\psi$ within $|y|<1$ in Au+Au collisions at $\sqrt{s_\mathrm{NN}} = 54.4$ GeV are shown in Fig.~\ref{fig-2} as a function of \pT\ for different centrality classes. The invariant yield is extracted as:
\begin{equation}
\setlength{\abovedisplayskip}{6pt}
\setlength{\belowdisplayskip}{6pt}
\begin{aligned}
B_{J/\psi \to e^{+}e^{-}}\frac{d^{2}N_{J/\psi}}{2\pi p_{\rm{T}}dp_{\rm{T}}dy} = \frac{\Delta N_{J/\psi}}{2\pi p_{\rm{T}}\Delta p_{\rm{T}} \Delta y  N_{\rm{MB}}  \epsilon_{\rm{total}}},
\end{aligned}
\end{equation}
where $B_{J/\psi \to e^{+}e^{-}}$ is the branching ratio of $J/\psi$ decaying into an $e^{+}e^{-}$ pair, $\Delta N_{J/\psi}$ is the raw $J/\psi$ count in each $p_{\rm{T}}$ bin for the considered centrality class, $p_{\rm{T}}$ is the bin center, $\Delta p_{\rm{T}}$ is the $p_{\rm{T}}$ bin width, $\Delta y$ is the rapidity coverage, from -1 to 1, $N_{\rm{MB}}$ stands for the number of MB events in the considered centrality class used in this analysis, and $\epsilon_{total}$ is the total $J/\psi$ efficiency.

Data points are placed at \pT\ values whose yields are equal to the average yields of the bins \cite{LAFFERTY1995541}. The \pT\ positions are determined by fitting the differential yields iteratively with an empirical function: 
\begin{equation}
\setlength{\abovedisplayskip}{6pt}
\setlength{\belowdisplayskip}{6pt}
\begin{aligned}
f(p_{\rm{T}}) = A \times p_{\rm{T}} \times (1+p_{\rm{T}}^{2}/B^{2})^{-C},
\end{aligned}
\label{empirical function}
\end{equation}
where $A$, $B$, and $C$ are free parameters. The horizontal bar around each data point indicates the bin width.

The nuclear modification factor ($R_{\rm{AA}}$) is used to quantify the modification to the $J/\psi$ production and is defined as:
\begin{equation}
\setlength{\abovedisplayskip}{6pt}
\setlength{\belowdisplayskip}{6pt}
\begin{aligned}
R_{\rm{AA}} = \frac{1}{\left\langle N_{\rm{coll}}\right\rangle/\sigma^{\rm{inelastic}}_{pp}}\frac{d^{2}N_{\rm{AA}}/dp_{\rm{T}}dy}{d^{2}\sigma_{pp}/dp_{\rm{T}}dy},
\end{aligned}
\label{raa}
\end{equation}
where $d^{2}N_{\rm{AA}}/dp_{\rm{T}}dy$ is the $J/\psi$ yield in A+A collisions and $d^{2}\sigma_{pp}/dp_{\rm{T}}dy$ is the $J/\psi$ cross section in $p$+$p$ collisions. $\sigma^{\rm{inelastic}}_{pp}$ is the inelastic $p$+$p$ cross section. Since there are no experimental measurements for the inclusive $J/\psi$ production cross section in $p$+$p$ collisions at $\sqrt{s} = 54.4$ GeV, a data-driven method is used to derive it by parameterizing collision energy, rapidity and $p_{\rm{T}}$ dependence of inclusive $J/\psi$ production cross section from world-data of $p$+$p$ and $p$+A collisions ranging between $\sqrt{s_\mathrm{NN}} = 6.8-7000$ GeV~\cite{PhysRevC.93.024919}. Differences in the interpolated values from different parameterizations are taken as systematic uncertainties~\cite{PhysRevC.93.024919}. The resulting interpolated inclusive $J/\psi$ production cross section, $Br_{e^{+}e^{-}}\frac{d\sigma}{dy}|_{|y|<1}$, for 54.4 GeV $p$+$p$ collisions is 14.39 $\pm$ 1.57 nb.

\begin{figure}[htbp]
\centering
\includegraphics[width=0.9\columnwidth,clip]{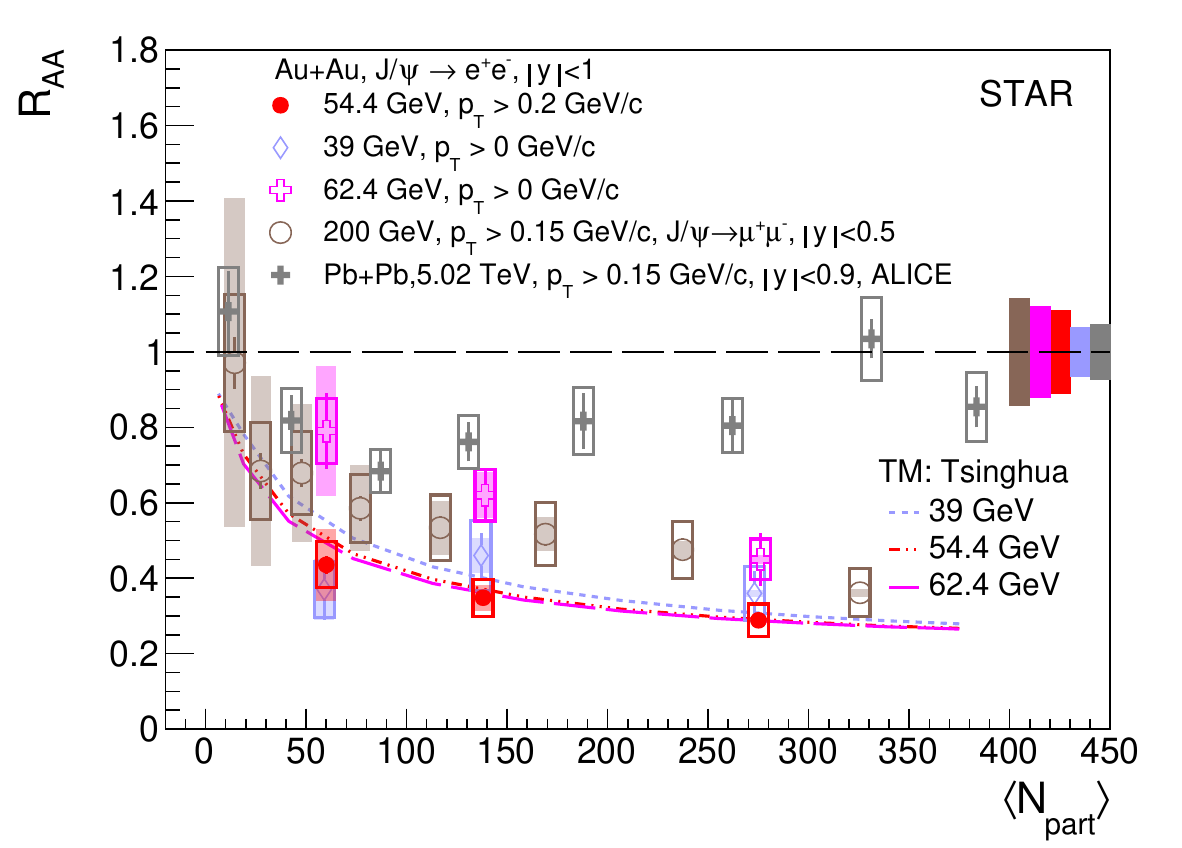}
\caption{The $R_{\rm{AA}}$ of inclusive $J/\psi$ at mid-rapidity as a function of $\langle N_{\rm{part}} \rangle$ in heavy-ion collisions at collision energies from 39 GeV to 5.02 TeV \cite{201713,2019134917,ALICE:2023gco}. Theoretical calculations are shown as dashed lines for comparison~\cite{PhysRevC.82.064905}. The error bars represent the statistical uncertainties, while the boxes represent the systematic uncertainties. The shaded bands on the STAR data points indicate the uncertainties in $\langle N_{\rm{coll}}\rangle$, while such uncertainties are included in the open boxes for the Pb+Pb data points. The bands around unity indicate the uncertainties from the reference $J/\psi$ cross sections in $p$+$p$ collisions, and those for 62.4 GeV and below are highly correlated~\cite{PhysRevC.93.024919}.}
\label{fig-3}       
\end{figure}

The \pT-integrated $R_{\rm{AA}}$ of inclusive $J/\psi$ as a function of $\langle N_{\rm{part}}\rangle$ is shown in Fig.~\ref{fig-3} for Au+Au collisions at collision energies of 39, 54.4, 62.4, 200 and 5020 GeV~\cite{201713,2019134917,ALICE:2023gco}. The newly measured $J/\psi$ $R_{\rm{AA}}$ at 54.4 GeV is consistent with previous results at RHIC within uncertainties, indicating no significant collision energy dependence of $R_{\rm{AA}}$ up to 200 GeV, while the precision is significantly improved. Taking into account statistical and systematic uncertainties from Au+Au measurements, the significance of the deviation in $R_{\rm{AA}}$ between 54.4 and 62.4 GeV varies between 1.6$\sigma$ and 2.5$\sigma$ depending on $\langle N_{\rm{part}}\rangle$. The distributions also hint at an increasing suppression from peripheral to central collisions at RHIC energies, consistent with increasing hot medium effects. On the other hand, the $J/\psi$ $R_{\rm{AA}}$ at the LHC seems to increase towards central collisions and is larger than at RHIC energies, which is attributed to a larger regeneration contribution at higher energies and more central collisions. Transport model calculations \cite{PhysRevC.105.064907} from the Tsinghua group for 39, 54.4 and 62.4 GeV are shown as dashed curves in Fig.~\ref{fig-3}, which predict very little difference among different energies as observed in data.

\begin{figure}[htbp]
\centering
\includegraphics[width=0.9\columnwidth,clip]{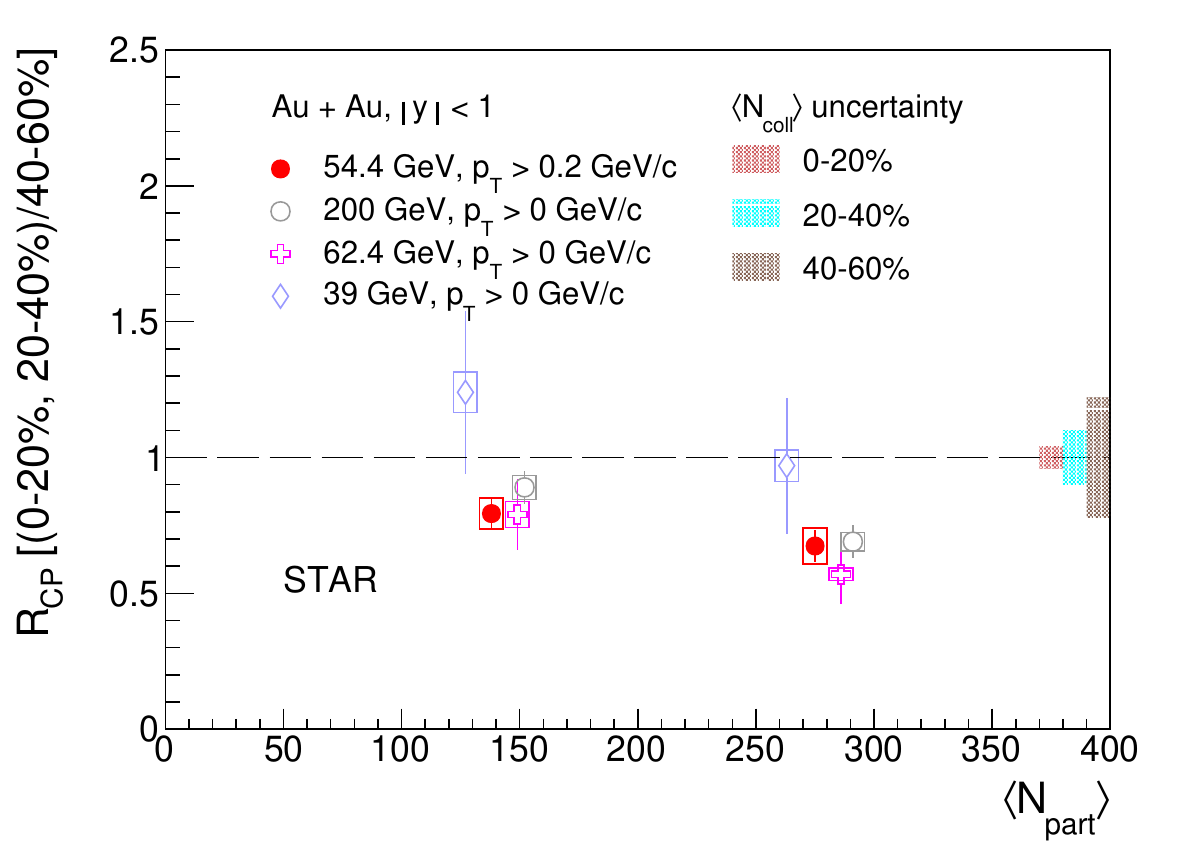}
\caption{The $R_{\rm{CP}}$ of inclusive $J/\psi$ at mid-rapidity as a function of $\langle N_{\rm{part}} \rangle$ in Au+Au collisions at collision energies from 39 GeV to 200 GeV \cite{201713}. The error bars represent the statistical uncertainties, while the boxes represent the systematic uncertainties. The shaded bands at unity indicate the uncertainties in $\langle N_{\rm{coll}}\rangle$ that apply to all energies.}
\label{fig-Rcp}       
\end{figure}

Unlike $R_{\rm{AA}}$, another nuclear modification factor, $R_{\rm{CP}}$, can be used to reflect the relative suppression between central and peripheral collisions. The $R_{\rm{CP}}$ is defined as a ratio of the $J/\psi$ yield in central collisions over that in peripheral collisions: 
\begin{equation}
\setlength{\abovedisplayskip}{6pt}
\setlength{\belowdisplayskip}{6pt}
\begin{aligned}
R_{\rm{CP}} = \frac{\frac{dN_{\rm{AA}}/dy}{\left\langle N_{\rm{coll}}\right\rangle}\enspace(\text{Central})}{\frac{dN_{AA}/dy}{\left\langle N_{\rm{coll}}\right\rangle}\enspace(\text{Peripheral})},
\end{aligned}
\label{raa}
\end{equation}
where $dN_{\rm{AA}}/dy$ is the \pT-integrated $J/\psi$ yield in a certain centrality class. One advantage of using $R_{\rm{CP}}$ is that it does not use the $J/\psi$ yield from $p$+$p$ collisions as the reference, which can only be interpolated from parameterization of world data for 39, 54.4 and 62.4 GeV. Furthermore, correlated systematic uncertainties in $J/\psi$ yield measurement largely cancel in the $R_{\rm{CP}}$. In this analysis, the peripheral 40-60\% centrality bin is used as the reference. Figure \ref{fig-Rcp} shows the centrality dependence of $R_{\rm{CP}}$ in Au+Au collisions at RHIC. The boxes on the data points represent systematic uncertainties, mainly from signal extraction, while uncertainties from tracking and electron identification mostly cancel. The remaining uncertainties in $\langle N_{\rm{coll}}\rangle$ after cancellation are indicated as shaded bands at unity. A relative suppression of $J/\psi$ yield in central collisions compared to that in peripheral collisions is observed at $\sqrt{s_\mathrm{NN}} = 54.4$ GeV. No significant collision energy dependence of $R_{\rm{CP}}$ has been observed at RHIC energies within uncertainties.

\begin{figure*}[htbp]
\centering
\includegraphics[width=0.9\textwidth,clip]{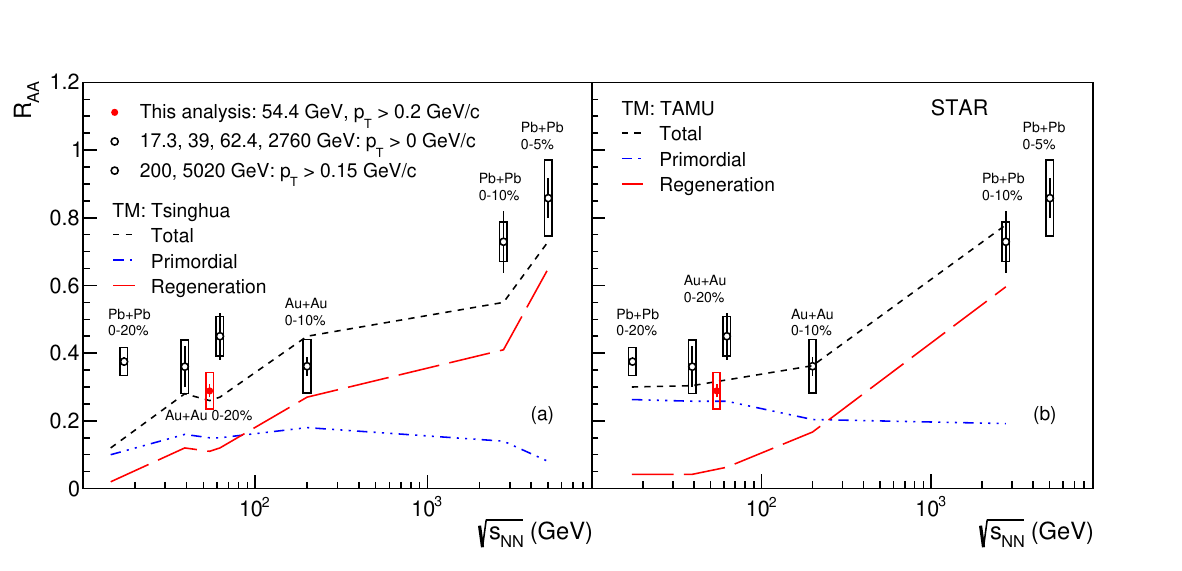}
\caption{The $R_{\rm{AA}}$ of $J/\psi$ as a function of collision energy in central collisions \cite{Kluberg:2005yh,200028,2014314,201713,2019134917,ALICE:2023gco}, in comparison with two transport model calculations from the Tsinghua group \cite{PhysRevC.105.064907} (left) and the TAMU group \cite{PhysRevC.82.064905} (right). The error bars represent the statistical uncertainties and the boxes represent the systematic uncertainties, including those from heavy-ion measurements, $p$+$p$ baseline and $\langle N_{\rm{coll}}\rangle$. The transport model calculations are shown as dashed lines for the total $J/\psi$ $R_{\rm{AA}}$, as dash-dot-tripled lines for the suppressed primordial production and long dash lines for the regeneration.}
\label{fig-4}       
\end{figure*}

Figure \ref{fig-4} shows the collision energy dependence of $J/\psi$ $R_{\rm{AA}}$ in central heavy-ion collisions of different species. 
No significant energy dependence is seen within uncertainties between 17.3 and 200 GeV in central collisions. Two transport model calculations of the collision energy dependence of inclusive $J/\psi$ $R_{\rm{AA}}$, displayed by the same line style, are shown on the left panel from the Tsinghua group \cite{PhysRevC.105.064907} and the right panel from the TAMU group \cite{PhysRevC.82.064905}. Blue dash-dot-tripled lines represent the suppression of primordial production due to CNM effects and dissociation in the QGP medium, while the red long dash lines denote the regeneration contribution. The 
final $J/\psi$ $R_{\rm{AA}}$ calculations from the two groups, considering both the suppression and regeneration, are shown as the black dash lines in the two panels of Fig.~\ref{fig-4}. The theoretical calculations, starting from 39 GeV, are consistent with the observed energy dependence of $J/\psi$ $R_{\rm{AA}}$, indicating that the $J/\psi$ production in high-energy heavy-ion collisions is an interplay of dissociation in the QGP medium, regeneration, and CNM effects. At 17.3 GeV, the transport model calculation from the Tsinghua group underestimates the experimental measurements.

\begin{figure*}[htbp]
\centering
\includegraphics[width=0.9\textwidth,clip]{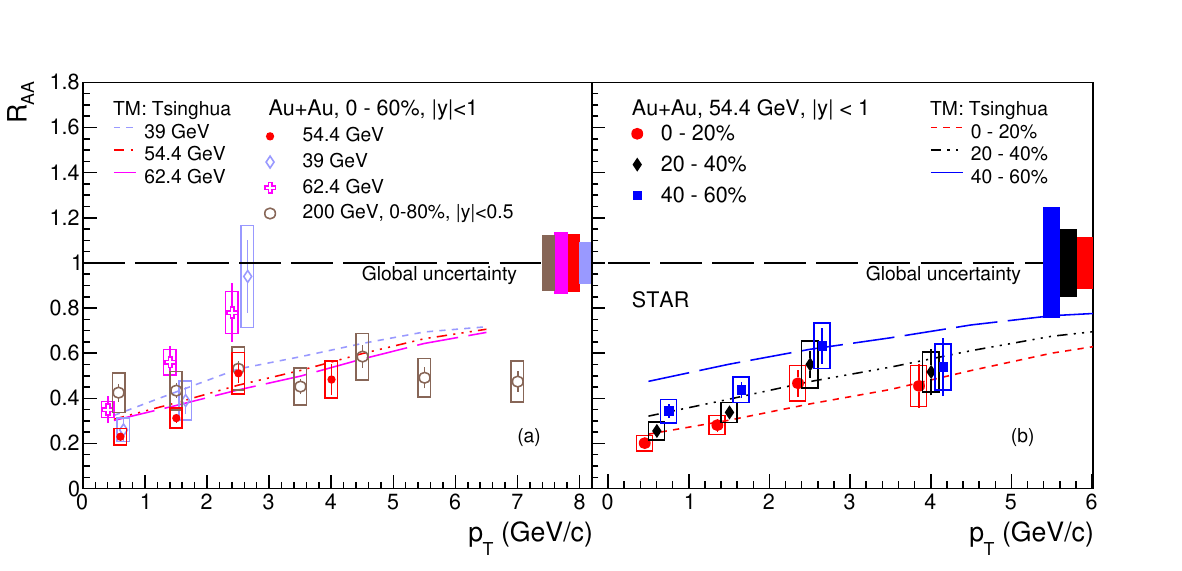}
\caption{$J/\psi$ $R_{\rm{AA}}$ as a function of \pT\ in the 0-60\% centrality class at different collision energies (left)~\cite{201713,2019134917} and for different centrality classes at 54.4 GeV (right). Theoretical calculations are shown as dashed lines for comparison~\cite{PhysRevC.82.064905}. The error bars represent the statistical uncertainties and the boxes represent the systematic uncertainties. The bands at unity show the relative uncertainties in the $p$+$p$ baseline and $\langle N_{\rm{coll}}\rangle$.}
\label{fig-5}       
\end{figure*}

Figure \ref{fig-5} shows $J/\psi$ $R_{\rm{AA}}$ as a function of \pT\ for the 0-60\% centrality class at different energies at RHIC~\cite{201713,2019134917} and for different centrality classes at 54.4 GeV. The \pT\ dependence seems to be flatter at $\sqrt{s_\mathrm{NN}} = 200$ GeV compared to lower energies, which could be due to a larger regeneration contribution at 200 GeV and larger nuclear absorption at lower energies at low \pT. The \pT\ dependence of $R_{\rm{AA}}$ also exhibits a similar distribution at 39, 54.4, and 62.4 GeV within uncertainties. 
In Fig. \ref{fig-5}(b), a slightly larger yield suppression is observed towards central collisions compared to that for peripheral collisions at $\sqrt{s_\mathrm{NN}} = 54.4$ GeV, while the \pT\ dependence is similar across different centrality classes. Transport calculations from the Tsinghua group, shown for comparison, are seen to overestimate \raa\ below 2 GeV/$c$ for 20-40\% and 40-60\% centrality classes at 54.4 GeV.  

\begin{figure}[htbp]
\centering
\includegraphics[width=0.9\columnwidth,clip]{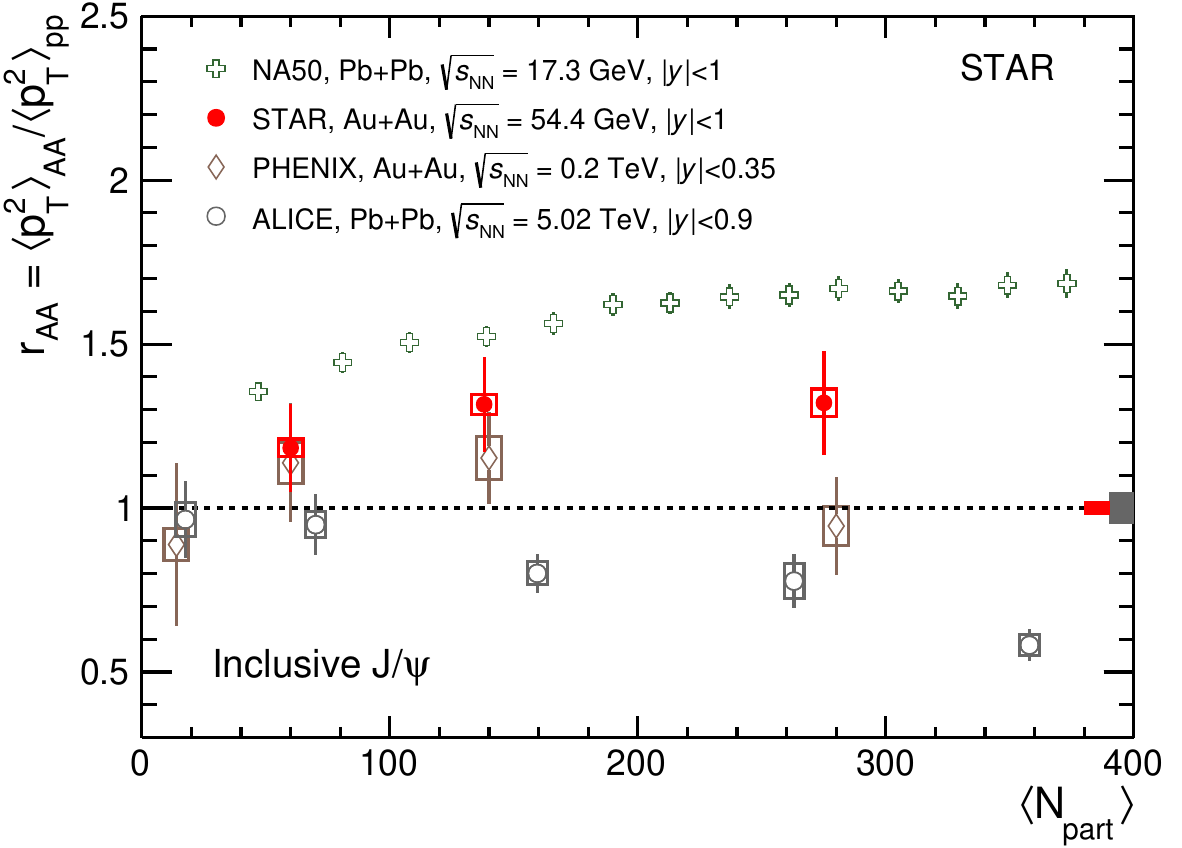}
\caption{The inclusive $J/\psi$ $r_{\rm{AA}}$ as a function of $\left\langle N_{\rm{part}} \right\rangle$ in different collision systems \cite{ALICE:2015nvt,PhysRevLett.98.232301,ABREU200185,PhysRevD.85.092004,PhysRevLett.101.122301,2020135434} at mid-rapidity. The error bars represent the statistical uncertainties and the boxes represent the systematic uncertainties. The bands at unity show the global uncertainty.}
\label{fig-6}
\end{figure}

The production of $J/\psi$ in high-energy heavy-ion collisions is an interplay of different hot and cold medium effects, while the $J/\psi$ \pT\ spectra shapes are crucial for studying these effects individually since different effects may dominate in different \pT\ regions. The second moment of the \pT\ distribution (\pTtwo) can facilitate the comparison of the measured $J/\psi$ \pT\ spectra shapes in different centrality bins at $\sqrt{s_\mathrm{NN}} = 54.4$ GeV as well as at different collision energies. The resulting \pTtwo\ in different centrality classes of Au+Au collisions at $\sqrt{s_\mathrm{NN}} = 54.4$ GeV are displayed in Table \ref{table3}, and no significant centrality dependence is seen. 
\begin{table}[htbp]
\centering
\begin{tabular}{|c@{\hspace{0.26cm}} c@{\hspace{0.26cm}} c@{\hspace{0.26cm}} c|}
\hline
    \multirow{2}{*}{}   & \multicolumn{3}{c|}{Au+Au, $\sqrt{s_\mathrm{NN}} = 54.4$ GeV}  \\
   &  0-20\% &  20-40\% & 40-60\%  \\
\hline
    $\langle p^{2}_{\rm{T}} \rangle$ &{\small$3.34\pm0.40\pm0.11$} & {\small$3.33\pm0.37\pm0.08$} & {\small$2.99\pm0.34\pm0.07$} \\
\hline    
\end{tabular}
\caption{Inclusive $J/\psi$ $\left\langle p^{2}_{\rm{T}} \right\rangle$ at mid-rapidity in Au+Au collisions at $\sqrt{s_\mathrm{NN}} = 54.4$ GeV for different centrality classes. The first and second uncertainties are statistical and systematic uncertainties, respectively.}\label{table3}
\label{table3}
\end{table}
The \pTtwo\ in $p$+$p$ collisions at $\sqrt{s} = 54.4$ GeV is derived to be $2.53\pm0.05$ $(\rm{GeV}/c)^{2}$~\cite{PhysRevC.93.024919}, where 0.05 is the total uncertainty, lower than in Au+Au collisions.  This is consistent with the observed \pT\ dependence of $R_{\rm{AA}}$, which increases towards higher \pT\ as shown in Fig. \ref{fig-5}. To quantify the change in \pTtwo, $r_{\rm{AA}}$ is used. It is defined as the ratio between \pTtwo$_{\rm{AA}}$ and \pTtwo$_{pp}$, and is shown as a function of $\langle N_{\rm{part}} \rangle$ in Fig. \ref{fig-6} for heavy-ion collisions with the collision energy ranging from 17.3 GeV to 5.02 TeV~\cite{ALICE:2015nvt,PhysRevLett.98.232301,ABREU200185,PhysRevD.85.092004,PhysRevLett.101.122301,2020135434}. The $r_{\rm{AA}}$ at $\sqrt{s_\mathrm{NN}} = 54.4$ GeV shows a flat distribution against centrality and follows the trend of the collision energy dependence. On the other hand, $r_{\rm{AA}}$ decreases towards central collisions at 5.02 TeV, due to increased regeneration contribution at low \pT.

\section{Summary}
We report on the measurements of inclusive $J/\psi$ production at mid-rapidity ($|y|<1$) in Au+Au collisions at $\sqrt{s_\mathrm{NN}} = 54.4$ GeV by the STAR experiment at RHIC. The newly measured $J/\psi$ $R_{\rm{AA}}$ at 54.4 GeV is compatible with previous measurements at 39 and 62.4 GeV with improved precision. A significant suppression of the $J/\psi$ yield is observed in the 0-60\% centrality class compared to the $p$+$p$ baseline. Hints of larger suppression towards central collisions and low \pT\ are seen, and no significant energy dependence of $J/\psi$ $R_{\rm{AA}}$ in central collisions from 17.3 to 200 GeV is observed within uncertainties. The $\langle N_{\rm{part}} \rangle$ dependence of $r_{\rm{AA}}$ at $\sqrt{s_\mathrm{NN}} = 54.4$ GeV shows a flat distribution and follows the trend of the collision energy dependence. Transport model calculations incorporating dissociation and regeneration in the QGP and the CNM effects can qualitatively describe the data. These results will provide additional constraints on understanding the interplay between different hot and cold nuclear matter effects, which are essential for probing the QGP properties with quarkonia.  

\section*{Acknowledgements}
We thank the RHIC Operations Group and SDCC at BNL, the NERSC Center at LBNL, and the Open Science Grid consortium for providing resources and support.  This work was supported in part by the Office of Nuclear Physics within the U.S. DOE Office of Science, the U.S. National Science Foundation, National Natural Science Foundation of China, Chinese Academy of Science, the Ministry of Science and Technology of China and the Chinese Ministry of Education, NSTC Taipei, the National Research Foundation of Korea, Czech Science Foundation and Ministry of Education, Youth and Sports of the Czech Republic, Hungarian National Research, Development and Innovation Office, New National Excellency Programme of the Hungarian Ministry of Human Capacities, Department of Atomic Energy and Department of Science and Technology of the Government of India, the National Science Centre and WUT ID-UB of Poland, German Bundesministerium f\"ur Bildung, Wissenschaft, Forschung and Technologie (BMBF), Helmholtz Association, Ministry of Education, Culture, Sports, Science, and Technology (MEXT), Japan Society for the Promotion of Science (JSPS), and Agencia Nacional de Investigacion y Desarrollo de Chile (ANID), Chile.
\label{sec4}




\bibliographystyle{elsarticle-num-names} 
\bibliography{reference.bib}

@article{STAR:2012wnc,
    author = "Adamczyk, L. and others",
    collaboration = "STAR",
    title = "{$J/\psi$ production at high transverse momenta in $p+p$ and Au+Au collisions at $\sqrt{s_{NN}} = 200$ GeV}",
    eprint = "1208.2736",
    archivePrefix = "arXiv",
    primaryClass = "nucl-ex",
    doi = "10.1016/j.physletb.2013.04.010",
    journal = "Phys. Lett. B",
    volume = "722",
    pages = "55--62",
    year = "2013"
}

@article{STAR:2013eve,
    author = "Adamczyk, L. and others",
    collaboration = "STAR",
    title = "{$J/\psi$ production at low $p_T$ in Au + Au and Cu + Cu collisions at $\sqrt{s_{NN}}=200$ GeV with the STAR detector}",
    eprint = "1310.3563",
    archivePrefix = "arXiv",
    primaryClass = "nucl-ex",
    doi = "10.1103/PhysRevC.90.024906",
    journal = "Phys. Rev. C",
    volume = "90",
    number = "2",
    pages = "024906",
    year = "2014"
}

@article{PHENIX:2012xtg,
    author = "Adare, A. and others",
    collaboration = "PHENIX",
    title = "{$J/\psi$ suppression at forward rapidity in Au+Au collisions at $\sqrt{s_{NN}}=39$ and 62.4 GeV}",
    eprint = "1208.2251",
    archivePrefix = "arXiv",
    primaryClass = "nucl-ex",
    doi = "10.1103/PhysRevC.86.064901",
    journal = "Phys. Rev. C",
    volume = "86",
    pages = "064901",
    year = "2012"
}

@article{ALICE:2022wpn,
    author = "Acharya, Shreyasi and others",
    collaboration = "ALICE",
    title = "{The ALICE experiment: a journey through QCD}",
    eprint = "2211.04384",
    archivePrefix = "arXiv",
    primaryClass = "nucl-ex",
    reportNumber = "CERN-EP-2022-227",
    doi = "10.1140/epjc/s10052-024-12935-y",
    journal = "Eur. Phys. J. C",
    volume = "84",
    number = "8",
    pages = "813",
    year = "2024"
}

@article{Heinz:2000bk,
    author = "Heinz, Ulrich W. and Jacob, Maurice",
    title = "{Evidence for a new state of matter: An Assessment of the results from the CERN lead beam program}",
    eprint = "nucl-th/0002042",
    archivePrefix = "arXiv",
    month = "1",
    year = "2000"
}

@article{ADAMS2005102,
title = {Experimental and theoretical challenges in the search for the quark–gluon plasma: The STAR Collaboration's critical assessment of the evidence from RHIC collisions},
journal = {Nuclear Physics A},
volume = {757},
number = {1},
pages = {102-183},
year = {2005},
note = {First Three Years of Operation of RHIC},
issn = {0375-9474},
doi = {https://doi.org/10.1016/j.nuclphysa.2005.03.085},
author = {J. Adams and others}
}

@article{ARSENE20051,
title = {Quark–gluon plasma and color glass condensate at RHIC? The perspective from the BRAHMS experiment},
journal = {Nuclear Physics A},
volume = {757},
number = {1},
pages = {1-27},
year = {2005},
note = {First Three Years of Operation of RHIC},
issn = {0375-9474},
doi = {https://doi.org/10.1016/j.nuclphysa.2005.02.130},
author = {I. Arsene and I.G. Bearden and D. Beavis and others}
}

@article{PHOBOS:2004zne,
    author = "Back, B. B. and others",
    collaboration = "PHOBOS",
    title = "{The PHOBOS perspective on discoveries at RHIC}",
    eprint = "nucl-ex/0410022",
    archivePrefix = "arXiv",
    doi = "10.1016/j.nuclphysa.2005.03.084",
    journal = "Nucl. Phys. A",
    volume = "757",
    pages = "28--101",
    year = "2005"
}

@article{PHENIX:2004vcz,
    author = "Adcox, K. and others",
    collaboration = "PHENIX",
    title = "{Formation of dense partonic matter in relativistic nucleus-nucleus collisions at RHIC: Experimental evaluation by the PHENIX collaboration}",
    eprint = "nucl-ex/0410003",
    archivePrefix = "arXiv",
    doi = "10.1016/j.nuclphysa.2005.03.086",
    journal = "Nucl. Phys. A",
    volume = "757",
    pages = "184--283",
    year = "2005"
}

@article{MATSUI1986416,
title = "{$J/\ensuremath{\psi}$ suppression by quark-gluon plasma formation}",
journal = {Physics Letters B},
volume = {178},
number = {4},
pages = {416-422},
year = {1986},
issn = {0370-2693},
doi = {https://doi.org/10.1016/0370-2693(86)91404-8},
author = {T. Matsui and H. Satz}
}

@article{BRAUNMUNZINGER2000196,
title = "{(Non)thermal aspects of charmonium production and a new look at $J/\ensuremath{\psi}$ suppression}",
journal = {Physics Letters B},
volume = {490},
number = {3},
pages = {196-202},
year = {2000},
issn = {0370-2693},
doi = {https://doi.org/10.1016/S0370-2693(00)00991-6},
author = {P. Braun-Munzinger and J. Stachel}
}

@article{PhysRevLett.92.212301,
  title = "{In-Medium Effects on Charmonium Production in Heavy-Ion Collisions}",
  author = {Grandchamp, Lo\"{\i}c and Rapp, Ralf and Brown, Gerald E.},
  journal = {Phys. Rev. Lett.},
  volume = {92},
  issue = {21},
  pages = {212301},
  numpages = {4},
  year = {2004},
  month = {May},
  publisher = {American Physical Society},
  doi = {10.1103/PhysRevLett.92.212301},
  url = {https://link.aps.org/doi/10.1103/PhysRevLett.92.212301}
}

@article{PhysRevC.84.044911,
  title = "{Modeling of $J/\ensuremath{\psi}$ modifications in deuteron-nucleus collisions at high energies}",
  author = {Nagle, J. L. and Frawley, A. D. and Levy, L. A. Linden and Wysocki, M. G.},
  journal = {Phys. Rev. C},
  volume = {84},
  issue = {4},
  pages = {044911},
  numpages = {11},
  year = {2011},
  month = {Oct},
  publisher = {American Physical Society},
  doi = {10.1103/PhysRevC.84.044911},
  url = {https://link.aps.org/doi/10.1103/PhysRevC.84.044911}
}

@article{NAGLE199921,
title = "{Initial state energy loss dependence of $J/\ensuremath{\psi}$ and Drell–Yan in relativistic heavy ion collisions}",
journal = {Physics Letters B},
volume = {465},
number = {1},
pages = {21-26},
year = {1999},
issn = {0370-2693},
doi = {https://doi.org/10.1016/S0370-2693(99)00988-0},
author = {J.L. Nagle and M.J. Bennett}
}

@article{PhysRevD.11.3105,
  title = "{Production of hadrons at large transverse momentum at 200, 300, and 400 GeV}",
  author = {Cronin, J. W. and Frisch, H. J. and Shochet, M. J. and others},
  journal = {Phys. Rev. D},
  volume = {11},
  issue = {11},
  pages = {3105--3123},
  numpages = {0},
  year = {1975},
  month = {Jun},
  publisher = {American Physical Society},
  doi = {10.1103/PhysRevD.11.3105},
  url = {https://link.aps.org/doi/10.1103/PhysRevD.11.3105}
}

@article{VOGT2002539,
title = "{Are the $J/\psi$ and $\chi$($c$) A dependencies the same?}",
journal = {Nuclear Physics A},
volume = {700},
number = {1},
pages = {539-554},
year = {2002},
issn = {0375-9474},
doi = {https://doi.org/10.1016/S0375-9474(01)01313-6},
author = {R. Vogt}
}

@article{FERREIRO201598,
title = "{Excited charmonium suppression in proton–nucleus collisions as a consequence of comovers}",
journal = {Physics Letters B},
volume = {749},
pages = {98-103},
year = {2015},
issn = {0370-2693},
doi = {https://doi.org/10.1016/j.physletb.2015.07.066},
author = {E.G. Ferreiro}
}

@article{Kluberg:2005yh,
    author = "Kluberg, L.",
    editor = "Lourenco, C. and Satz, H.",
    title = "{20 years of $J/\ensuremath{\psi}$ suppression at the CERN SPS: Results from experiments NA38, NA51 and NA50}",
    doi = "10.1140/epjc/s2005-02245-6",
    journal = "Eur. Phys. J. C",
    volume = "43",
    pages = "145--156",
    year = "2005"
}

@article{200028,
title = "{Evidence for deconfinement of quarks and gluons from the $J/\ensuremath{\psi}$ suppression pattern measured in Pb-Pb collisions at the CERN-SPS}",
journal = {Physics Letters B},
volume = {477},
number = {1},
pages = {28-36},
year = {2000},
issn = {0370-2693},
doi = {https://doi.org/10.1016/S0370-2693(00)00237-9},
author = {M.C. Abreu and B. Alessandro and C. Alexa and others}
}

@article{2014314,
title = "{Centrality, rapidity and transverse momentum dependence of $J/\ensuremath{\psi}$ suppression in Pb–Pb collisions at $\sqrt{s_{NN}}$=2.76 TeV}",
journal = {Physics Letters B},
volume = {734},
pages = {314-327},
year = {2014},
issn = {0370-2693},
doi = {https://doi.org/10.1016/j.physletb.2014.05.064},
author = {B. Abelev and J. Adam and D. Adamová and others}
}

@article{201713,
title = "{Energy dependence of $J/\ensuremath{\psi}$ production in Au+Au collisions at $\sqrt{s_{\mathrm{NN}}}$=39,62.4 and 200 GeV}",
journal = {Physics Letters B},
volume = {771},
pages = {13-20},
year = {2017},
issn = {0370-2693},
doi = {https://doi.org/10.1016/j.physletb.2017.04.078},
author = {L. Adamczyk and J.K. Adkins and G. Agakishiev and others}
}

@article{2019134917,
title = "{Measurement of inclusive $J/\ensuremath{\psi}$ suppression in Au+Au collisions at $\sqrt{s_{\mathrm{NN}}}$=200 GeV through the dimuon channel at STAR}",
journal = {Physics Letters B},
volume = {797},
pages = {134917},
year = {2019},
issn = {0370-2693},
doi = {https://doi.org/10.1016/j.physletb.2019.134917},
author = {J. Adam and L. Adamczyk and J.R. Adams and others}
}

@article{BAI2021121769,
title = "{Quarkonium measurements in nucleus-nucleus collisions with ALICE}",
journal = {Nuclear Physics A},
volume = {1005},
pages = {121769},
year = {2021},
note = {The 28th International Conference on Ultra-relativistic Nucleus-Nucleus Collisions: Quark Matter 2019},
issn = {0375-9474},
doi = {https://doi.org/10.1016/j.nuclphysa.2020.121769},
author = {Xiaozhi Bai}
}

@article{LLOPE2012S110,
title = "{Multigap RPCs in the STAR experiment at RHIC}",
journal = {Nuclear Instruments and Methods in Physics Research Section A: Accelerators, Spectrometers, Detectors and Associated Equipment},
volume = {661},
pages = {S110-S113},
year = {2012},
note = {X. Workshop on Resistive Plate Chambers and Related Detectors (RPC 2010)},
issn = {0168-9002},
doi = {https://doi.org/10.1016/j.nima.2010.07.086},
author = {W.J. Llope}
}

@article{PhysRevC.93.024919,
  title = "{Systematic study of the experimental measurements on $J/\ensuremath{\psi}$ cross sections and kinematic distributions in $p+p$ collisions at different energies}",
  author = {Zha, Wangmei and Huang, Bingchu and Ma, Rongrong and others},
  journal = {Phys. Rev. C},
  volume = {93},
  issue = {2},
  pages = {024919},
  numpages = {7},
  year = {2016},
  month = {Feb},
  publisher = {American Physical Society},
  doi = {10.1103/PhysRevC.93.024919},
  url = {https://link.aps.org/doi/10.1103/PhysRevC.93.024919}
}

@article{PhysRevC.82.064905,
  title = "{Charmonium in medium: From correlators to experiment}",
  author = {Zhao, Xingbo and Rapp, Ralf},
  journal = {Phys. Rev. C},
  volume = {82},
  issue = {6},
  pages = {064905},
  numpages = {16},
  year = {2010},
  month = {Dec},
  publisher = {American Physical Society},
  doi = {10.1103/PhysRevC.82.064905},
  url = {https://link.aps.org/doi/10.1103/PhysRevC.82.064905}
}

@article{PhysRevC.53.3051,
  title = {J/\ensuremath{\psi} suppression in an equilibrating parton plasma},
  author = {Xu, Xiao-Ming and Kharzeev, D. and Satz, H. and others},
  journal = {Phys. Rev. C},
  volume = {53},
  issue = {6},
  pages = {3051--3056},
  numpages = {0},
  year = {1996},
  month = {Jun},
  publisher = {American Physical Society},
  doi = {10.1103/PhysRevC.53.3051},
  url = {https://link.aps.org/doi/10.1103/PhysRevC.53.3051}
}

@article{PhysRevD.100.014008,
  title = {Quarkonium inside the quark-gluon plasma: Diffusion, dissociation, recombination, and energy loss},
  author = {Yao, Xiaojun and M\"uller, Berndt},
  journal = {Phys. Rev. D},
  volume = {100},
  issue = {1},
  pages = {014008},
  numpages = {16},
  year = {2019},
  month = {Jul},
  publisher = {American Physical Society},
  doi = {10.1103/PhysRevD.100.014008},
  url = {https://link.aps.org/doi/10.1103/PhysRevD.100.014008}
}

@article{PhysRevC.87.044905,
  title = {High transverse momentum quarkonium production and dissociation in heavy ion collisions},
  author = {Sharma, Rishi and Vitev, Ivan},
  journal = {Phys. Rev. C},
  volume = {87},
  issue = {4},
  pages = {044905},
  numpages = {23},
  year = {2013},
  month = {Apr},
  publisher = {American Physical Society},
  doi = {10.1103/PhysRevC.87.044905},
  url = {https://link.aps.org/doi/10.1103/PhysRevC.87.044905}
}

@article{PhysRevLett.98.232301,
  title = {$\mathrm{J}/\ensuremath{\psi}$ Production versus Centrality, Transverse Momentum, and Rapidity in $\mathrm{Au}+\mathrm{Au}$ Collisions at $\sqrt{{s}_{NN}}=200\text{ }\text{ }\mathrm{GeV}$},
  author = {Adare, A. and Afanasiev, S. and Aidala, C. and others},
  collaboration = {PHENIX Collaboration},
  journal = {Phys. Rev. Lett.},
  volume = {98},
  issue = {23},
  pages = {232301},
  numpages = {6},
  year = {2007},
  month = {Jun},
  publisher = {American Physical Society},
  doi = {10.1103/PhysRevLett.98.232301},
  url = {https://link.aps.org/doi/10.1103/PhysRevLett.98.232301}
}

@article{PhysRevLett.97.232301,
  title = {$\mathrm{J}/\ensuremath{\psi}$ Production in Quark-Gluon Plasma},
  author = {Yan, Li and Zhuang, Pengfei and Xu, Nu},
  journal = {Phys. Rev. Lett.},
  volume = {97},
  issue = {23},
  pages = {232301},
  numpages = {4},
  year = {2006},
  month = {Dec},
  publisher = {American Physical Society},
  doi = {10.1103/PhysRevLett.97.232301},
  url = {https://link.aps.org/doi/10.1103/PhysRevLett.97.232301}
}

@article{PhysRevC.89.054911,
  title = {Medium effects on charmonium production at ultrarelativistic energies available at the CERN Large Hadron Collider},
  author = {Zhou, Kai and Xu, Nu and Xu, Zhe and others},
  journal = {Phys. Rev. C},
  volume = {89},
  issue = {5},
  pages = {054911},
  numpages = {11},
  year = {2014},
  month = {May},
  publisher = {American Physical Society},
  doi = {10.1103/PhysRevC.89.054911},
  url = {https://link.aps.org/doi/10.1103/PhysRevC.89.054911}
}

@article{ZHAO2011114,
title = {Medium modifications and production of charmonia at LHC},
journal = {Nuclear Physics A},
volume = {859},
number = {1},
pages = {114-125},
year = {2011},
issn = {0375-9474},
doi = {https://doi.org/10.1016/j.nuclphysa.2011.05.001},
author = {Xingbo Zhao and Ralf Rapp}
}

@article{Andronic:2017pug,
    author = "Andronic, Anton and Braun-Munzinger, Peter and Redlich, Krzysztof and others",
    title = "{Decoding the phase structure of QCD via particle production at high energy}",
    eprint = "1710.09425",
    archivePrefix = "arXiv",
    primaryClass = "nucl-th",
    doi = "10.1038/s41586-018-0491-6",
    journal = "Nature",
    volume = "561",
    number = "7723",
    pages = "321--330",
    year = "2018"
}

@article{Judd:2018zbg,
    author = "Judd, E. G. and others",
    title = "{The evolution of the STAR Trigger System}",
    doi = "10.1016/j.nima.2018.03.070",
    journal = "Nucl. Instrum. Meth. A",
    volume = "902",
    pages = "228--237",
    year = "2018"
}

@article{Adler:2000bd,
    author = "Adler, Clemens and Denisov, Alexei and Garcia, Edmundo and others",
    title = "{The RHIC zero degree calorimeter}",
    eprint = "nucl-ex/0008005",
    archivePrefix = "arXiv",
    doi = "10.1016/S0168-9002(01)00627-1",
    journal = "Nucl. Instrum. Meth. A",
    volume = "470",
    pages = "488--499",
    year = "2001"
}

@article{LLOPE201423,
title = {The STAR Vertex Position Detector},
journal = {Nuclear Instruments and Methods in Physics Research Section A: Accelerators, Spectrometers, Detectors and Associated Equipment},
volume = {759},
pages = {23-28},
year = {2014},
issn = {0168-9002},
doi = {https://doi.org/10.1016/j.nima.2014.04.080},
author = {W.J. Llope and J. Zhou and T. Nussbaum and others}
}

@article{Anderson:2003ur,
    author = "Anderson, M. and others",
    title = "{The Star time projection chamber: A Unique tool for studying high multiplicity events at RHIC}",
    eprint = "nucl-ex/0301015",
    archivePrefix = "arXiv",
    doi = "10.1016/S0168-9002(02)01964-2",
    journal = "Nucl. Instrum. Meth. A",
    volume = "499",
    pages = "659--678",
    year = "2003"
}

@article{STAR:2002ymp,
    author = "Beddo, M. and others",
    collaboration = "STAR",
    title = "{The STAR barrel electromagnetic calorimeter}",
    doi = "10.1016/S0168-9002(02)01970-8",
    journal = "Nucl. Instrum. Meth. A",
    volume = "499",
    pages = "725--739",
    year = "2003"
}

@article{PhysRevC.79.034909,
  title = {Systematic measurements of identified particle spectra in $\mathit{pp}$, $d+\mathrm{Au}$, and $\mathrm{Au}+\mathrm{Au}$ collisions at the STAR detector},
  author = {Abelev, B. I. and Aggarwal, M. M. and Ahammed, Z. and others},
  collaboration = {STAR Collaboration},
  journal = {Phys. Rev. C},
  volume = {79},
  issue = {3},
  pages = {034909},
  numpages = {58},
  year = {2009},
  month = {Mar},
  publisher = {American Physical Society},
  doi = {10.1103/PhysRevC.79.034909},
  url = {https://link.aps.org/doi/10.1103/PhysRevC.79.034909}
}

@article{PhysRevLett.113.022301,
  title = {Dielectron Mass Spectra from $\mathrm{Au}+\mathrm{Au}$ Collisions at $\sqrt{{s}_{NN}}=200\text{ }\text{ }\mathrm{GeV}$},
  author = {Adamczyk, L. and Adkins, J. K. and Agakishiev, G. and others},
  collaboration = {STAR Collaboration},
  journal = {Phys. Rev. Lett.},
  volume = {113},
  issue = {2},
  pages = {022301},
  numpages = {7},
  year = {2014},
  month = {Jul},
  publisher = {American Physical Society},
  doi = {10.1103/PhysRevLett.113.022301},
  url = {https://link.aps.org/doi/10.1103/PhysRevLett.113.022301}
}

@article{ALICE:2015nvt,
    author = "Adam, Jaroslav and others",
    collaboration = "ALICE",
    title = "{Inclusive, prompt and non-prompt $J/\psi$ production at mid-rapidity in Pb-Pb collisions at $\sqrt{s_{\rm NN}}$ = 2.76 TeV}",
    eprint = "1504.07151",
    archivePrefix = "arXiv",
    primaryClass = "nucl-ex",
    reportNumber = "CERN-PH-EP-2015-092",
    doi = "10.1007/JHEP07(2015)051",
    journal = "JHEP",
    volume = "07",
    pages = "051",
    year = "2015"
}

@article{ABREU200185,
title = {Transverse momentum distributions of $\mathrm{J}/\psi$, $\psi^\prime$, Drell–Yan and continuum dimuons produced in Pb–Pb interactions at the SPS},
journal = {Physics Letters B},
volume = {499},
number = {1},
pages = {85-96},
year = {2001},
issn = {0370-2693},
doi = {https://doi.org/10.1016/S0370-2693(01)00019-3},
author = {M.C Abreu and B Alessandro and C Alexa and others}
}

@article{PhysRevD.85.092004,
  title = {Ground and excited state charmonium production in $p+p$ collisions at $\sqrt{s}=200\text{ }\text{ }\mathrm{GeV}$},
  author = {Adare, A. and Afanasiev, S. and Aidala, C. and others},
  collaboration = {PHENIX Collaboration},
  journal = {Phys. Rev. D},
  volume = {85},
  issue = {9},
  pages = {092004},
  numpages = {27},
  year = {2012},
  month = {May},
  publisher = {American Physical Society},
  doi = {10.1103/PhysRevD.85.092004},
  url = {https://link.aps.org/doi/10.1103/PhysRevD.85.092004}
}

@article{PhysRevLett.101.122301,
  title = {$\mathrm{J}/\ensuremath{\psi}$ Production in $\sqrt{{s}_{NN}}=200\text{ }\text{ }\mathrm{GeV}$ $\mathrm{Cu}+\mathrm{Cu}$ Collisions},
  author = {Adare, A. and Afanasiev, S. and Aidala, C. and others},
  collaboration = {PHENIX Collaboration},
  journal = {Phys. Rev. Lett.},
  volume = {101},
  issue = {12},
  pages = {122301},
  numpages = {6},
  year = {2008},
  month = {Sep},
  publisher = {American Physical Society},
  doi = {10.1103/PhysRevLett.101.122301},
  url = {https://link.aps.org/doi/10.1103/PhysRevLett.101.122301}
}

@article{2020135434,
title = {Centrality and transverse momentum dependence of inclusive $\mathrm{J}/\psi$ production at midrapidity in $\mathrm{{Pb}}$+$\mathrm{Pb}$ collisions at $\sqrt{s_{\mathrm{NN}}}$=5.02 TeV},
journal = {Physics Letters B},
volume = {805},
pages = {135434},
year = {2020},
issn = {0370-2693},
doi = {https://doi.org/10.1016/j.physletb.2020.135434},
author = {S. Acharya and D. Adamová and A. Adler and others}
}

@article{BICHSEL2006154,
title = {A method to improve tracking and particle identification in TPCs and silicon detectors},
journal = {Nuclear Instruments and Methods in Physics Research Section A: Accelerators, Spectrometers, Detectors and Associated Equipment},
volume = {562},
number = {1},
pages = {154-197},
year = {2006},
issn = {0168-9002},
doi = {https://doi.org/10.1016/j.nima.2006.03.009},
author = {Hans Bichsel}
}

@article{PhysRevC.81.064904,
  author = "Abelev, B. I. and others",
    collaboration = "STAR",
    title = "{Inclusive $\pi^0$, $\eta$, and direct photon production at high transverse momentum in $p+p$ and $d+$Au collisions at $\sqrt{s_{NN}}=200$ GeV}",
    eprint = "0912.3838",
    archivePrefix = "arXiv",
    primaryClass = "hep-ex",
    doi = "10.1103/PhysRevC.81.064904",
    journal = "Phys. Rev. C",
    volume = "81",
    pages = "064904",
    year = "2010"
}

@article{PhysRevC.97.044910,
  title = {Coherent $\mathrm{J}/\ensuremath{\psi}$ photoproduction in hadronic heavy-ion collisions},
  author = {Zha, W. and Klein, S. R. and Ma, R. and Ruan, L. and Todoroki, T. and Tang, Z. and Xu, Z. and Yang, C. and Yang, Q. and Yang, S.},
  journal = {Phys. Rev. C},
  volume = {97},
  issue = {4},
  pages = {044910},
  numpages = {6},
  year = {2018},
  month = {Apr},
  publisher = {American Physical Society},
  doi = {10.1103/PhysRevC.97.044910},
  url = {https://link.aps.org/doi/10.1103/PhysRevC.97.044910}
}

@article{PhysRevLett.123.132302,
  author = "Adam, J. and others",
    collaboration = "STAR",
    title = "{Observation of excess $J/\psi$ yield at very low transverse momenta in Au+Au collisions at $\sqrt{s_{\rm{NN}}} =$ 200 GeV and U+U collisions at $\sqrt{s_{\rm{NN}}} =$ 193 GeV}",
    eprint = "1904.11658",
    archivePrefix = "arXiv",
    primaryClass = "hep-ex",
    doi = "10.1103/PhysRevLett.123.132302",
    journal = "Phys. Rev. Lett.",
    volume = "123",
    number = "13",
    pages = "132302",
    year = "2019"
}

@article{PhysRevLett.116.222301,
   author = "Adam, Jaroslav and others",
    collaboration = "ALICE",
    title = "{Measurement of an excess in the yield of $J/\psi$ at very low $p_{\rm T}$ in Pb-Pb collisions at $\sqrt{s_{\rm NN}}$ = 2.76 TeV}",
    eprint = "1509.08802",
    archivePrefix = "arXiv",
    primaryClass = "nucl-ex",
    reportNumber = "CERN-PH-EP-2015-268",
    doi = "10.1103/PhysRevLett.116.222301",
    journal = "Phys. Rev. Lett.",
    volume = "116",
    number = "22",
    pages = "222301",
    year = "2016"
}

@article{PhysRevC.81.034911,
  author = "Adare, A. and others",
    collaboration = "PHENIX",
    title = "{Detailed measurement of the $e^+ e^-$ pair continuum in $p+p$ and Au+Au collisions at $\sqrt{s_{NN}} = 200$ GeV and implications for direct photon production}",
    eprint = "0912.0244",
    archivePrefix = "arXiv",
    primaryClass = "nucl-ex",
    doi = "10.1103/PhysRevC.81.034911",
    journal = "Phys. Rev. C",
    volume = "81",
    pages = "034911",
    year = "2010"
}

@article{PhysRevC.105.064907,
  title = {Effects of cold and hot nuclear matter on $\mathrm{J}/\ensuremath{\psi}$ production at energies selected for the beam energy scan at the BNL Relativistic Heavy Ion Collider},
  author = {Zhao, Jiaxing and Zhuang, Pengfei},
  journal = {Phys. Rev. C},
  volume = {105},
  issue = {6},
  pages = {064907},
  numpages = {8},
  year = {2022},
  month = {Jun},
  publisher = {American Physical Society},
  doi = {10.1103/PhysRevC.105.064907},
  url = {https://link.aps.org/doi/10.1103/PhysRevC.105.064907}
}

@article{LAFFERTY1995541,
title = {Where to stick your data points: The treatment of measurements within wide bins},
journal = {Nuclear Instruments and Methods in Physics Research Section A: Accelerators, Spectrometers, Detectors and Associated Equipment},
volume = {355},
number = {2},
pages = {541-547},
year = {1995},
issn = {0168-9002},
doi = {https://doi.org/10.1016/0168-9002(94)01112-5},
author = {G.D. Lafferty and T.R. Wyatt}
}

@article{ALICE:2023gco,
    author = "Acharya, Shreyasi and others",
    collaboration = "ALICE",
    title = "{Measurements of inclusive J/{\ensuremath{\psi}} production at midrapidity and forward rapidity in Pb{\textendash}Pb collisions at sNN = 5.02 TeV}",
    eprint = "2303.13361",
    archivePrefix = "arXiv",
    primaryClass = "nucl-ex",
    reportNumber = "CERN-EP-2023-054",
    doi = "10.1016/j.physletb.2024.138451",
    journal = "Phys. Lett. B",
    volume = "849",
    pages = "138451",
    year = "2024"
}

@article{PhysRevLett.130.202301,
  title = {Beam Energy Dependence of Triton Production and Yield Ratio (${N}_{t}\ifmmode\times\else\texttimes\fi{}{N}_{p}/{N}_{d}^{2}$) in $\mathrm{Au}+\mathrm{Au}$ Collisions at RHIC},
  collaboration = {STAR Collaboration},
  journal = {Phys. Rev. Lett.},
  volume = {130},
  issue = {20},
  pages = {202301},
  numpages = {9},
  year = {2023},
  month = {May},
  publisher = {American Physical Society},
  doi = {10.1103/PhysRevLett.130.202301},
  url = {https://link.aps.org/doi/10.1103/PhysRevLett.130.202301}
}

@article{PhysRevLett.33.1404,
  title = {Experimental Observation of a Heavy Particle $\mathrm{J}$},
  author = {Aubert, J. J. and Becker, U. and Biggs, P. J. and Burger, J. and Chen, M. and Everhart, G. and Goldhagen, P. and Leong, J. and McCorriston, T. and Rhoades, T. G. and Rohde, M. and Ting, Samuel C. C. and Wu, Sau Lan and Lee, Y. Y.},
  journal = {Phys. Rev. Lett.},
  volume = {33},
  issue = {23},
  pages = {1404--1406},
  numpages = {0},
  year = {1974},
  month = {Dec},
  publisher = {American Physical Society},
  doi = {10.1103/PhysRevLett.33.1404},
  url = {https://link.aps.org/doi/10.1103/PhysRevLett.33.1404}
}

@article{PhysRevLett.33.1406,
  title = {Discovery of a Narrow Resonance in ${e}^{+}{e}^{\ensuremath{-}}$ Annihilation},
  author = {Augustin, J. -E. and Boyarski, A. M. and Breidenbach, M. and Bulos, F. and Dakin, J. T. and Feldman, G. J. and Fischer, G. E. and Fryberger, D. and Hanson, G. and Jean-Marie, B. and Larsen, R. R. and L\"uth, V. and Lynch, H. L. and Lyon, D. and Morehouse, C. C. and Paterson, J. M. and Perl, M. L. and Richter, B. and Rapidis, P. and Schwitters, R. F. and Tanenbaum, W. M. and Vannucci, F. and Abrams, G. S. and Briggs, D. and Chinowsky, W. and Friedberg, C. E. and Goldhaber, G. and Hollebeek, R. J. and Kadyk, J. A. and Lulu, B. and Pierre, F. and Trilling, G. H. and Whitaker, J. S. and Wiss, J. and Zipse, J. E.},
  journal = {Phys. Rev. Lett.},
  volume = {33},
  issue = {23},
  pages = {1406--1408},
  numpages = {0},
  year = {1974},
  month = {Dec},
  publisher = {American Physical Society},
  doi = {10.1103/PhysRevLett.33.1406},
  url = {https://link.aps.org/doi/10.1103/PhysRevLett.33.1406}
}

@article{STAR:2018smh,
    author = "Adam, Jaroslav and others",
    collaboration = "STAR",
    title = "{$J/\psi$ production cross section and its dependence on charged-particle multiplicity in $p + p$ collisions at $\sqrt{s}$ = 200 GeV}",
    eprint = "1805.03745",
    archivePrefix = "arXiv",
    primaryClass = "hep-ex",
    doi = "10.1016/j.physletb.2018.09.029",
    journal = "Phys. Lett. B",
    volume = "786",
    pages = "87--93",
    year = "2018"
}

@article{PHENIX:2009ghc,
    author = "Adare, A. and others",
    collaboration = "PHENIX",
    title = "{Transverse momentum dependence of $J/\psi$ polarization at midrapidity in p+p collisions at $\sqrt{s}$ = 200 GeV}",
    eprint = "0912.2082",
    archivePrefix = "arXiv",
    primaryClass = "hep-ex",
    doi = "10.1103/PhysRevD.82.012001",
    journal = "Phys. Rev. D",
    volume = "82",
    pages = "012001",
    year = "2010"
}

@article{Smith:2022pro,
    author = "Smith, Krista",
    collaboration = "PHENIX",
    title = "{$J/\psi $ and $\psi (2S)$ Production in Small Systems with PHENIX}",
    eprint = "2212.08885",
    archivePrefix = "arXiv",
    primaryClass = "nucl-ex",
    doi = "10.5506/APhysPolBSupp.16.1-A73",
    journal = "Acta Phys. Polon. Supp.",
    volume = "16",
    number = "1",
    pages = "1--A73",
    year = "2023"
}

@article{Chen:2024aom,
    author = "Chen, Jinhui and others",
    title = "{Properties of the QCD matter: review of selected results from the relativistic heavy ion collider beam energy scan (RHIC BES) program}",
    eprint = "2407.02935",
    archivePrefix = "arXiv",
    primaryClass = "nucl-ex",
    doi = "10.1007/s41365-024-01591-2",
    journal = "Nucl. Sci. Tech.",
    volume = "35",
    number = "12",
    pages = "214",
    year = "2024"
}

@phdthesis{Skwarnicki:1986xj,
    author = "Skwarnicki, Tomasz",
    title = "{A study of the radiative CASCADE transitions between the Upsilon-Prime and Upsilon resonances}",
    reportNumber = "DESY-F31-86-02, DESY-F-31-86-02",
    school = "Cracow, INP",
    year = "1986"
}

@mastersthesis{Oreglia:1980cs,
    author = "Oreglia, M.",
    title = "{A Study of the Reactions $\psi^\prime \to \gamma \gamma \psi$}",
    reportNumber = "SLAC-0236, SLAC-236, UMI-81-08973, SLAC-R-0236, SLAC-R-236",
    type = "Ph.\uppercase{D}. thesis",
    month = "12",
    year = "1980"
}




\end{document}